\documentclass[
aip,jcp,
reprint
]{revtex4-1}

\usepackage{amsmath}
\usepackage{amsfonts}
\usepackage{url}
\usepackage{graphicx}
\usepackage{color}

\newcommand{\dd}[2]{\frac{d#1}{d#2}}

\newcommand{\bi}{\mathbf{i}}
\newcommand{\bj}{\mathbf{j}}
\newcommand{\bzero}{\mathbf{0}}
\newcommand{\bH}{\mathbf{H}}
\newcommand{\bT}{\mathbf{T}}
\newcommand{\bV}{\mathbf{V}}
\newcommand{\bI}{\mathbf{I}}
\newcommand{\bc}{\mathbf{c}}
\newcommand{\bn}{\mathbf{n}}
\newcommand{\bp}{\mathbf{p}}
\newcommand{\bk}{\mathbf{k}}
\newcommand{\br}{\mathbf{r}}

\newcommand{\bS}{\mathbf{S}}

\newcommand{\EHF}{E_{\mathrm{HF}}}
\newcommand{\hyperF}[2]{{_0}F_{1}\left(;#1;#2\right)}

\newcommand{\bra}{\langle}
\newcommand{\ket}{\rangle}

\newcommand{\fref}[1]{Fig.~\ref{#1}}
\newcommand{\eref}[1]{Eq.~\ref{#1}}
\newcommand{\sref}[1]{Sec.~\ref{#1}}

\DeclareMathOperator{\sign}{sign}

\begin{document}

\title{The sign problem and population dynamics in the full configuration interaction quantum Monte Carlo method}

\author{J.S.~Spencer}
\affiliation{Department of Materials, Imperial College London, Exhibition Road, London, SW7 2AZ, U.K.}
\affiliation{Department of Physics, Imperial College London, Exhibition Road, London, SW7 2AZ, U.K.}
\author{N.S.~Blunt}
\affiliation{Department of Physics, Imperial College London, Exhibition Road, London, SW7 2AZ, U.K.}
\author{W.M.C.~Foulkes}
\affiliation{Department of Physics, Imperial College London, Exhibition Road, London, SW7 2AZ, U.K.}

\begin{abstract}
The recently proposed full configuration interaction quantum Monte Carlo method allows access to essentially exact ground-state energies of systems of interacting fermions substantially larger than previously tractable without knowledge of the nodal structure of the ground-state wave function.  We investigate the nature of the sign problem in this method and how its severity depends on the system studied. We explain how cancelation of the positive and negative particles sampling the wave function ensures convergence to a stochastic representation of the many-fermion ground state and accounts for the characteristic population dynamics observed in simulations.
\end{abstract}

\maketitle

\section{Introduction}

% Ground state calculations
%  -> FCI method
%  -> FCIQMC method
One of the major goals of electronic structure methods is to produce accurate ground-state energies and properties of many-electron systems.\cite{Knowles2000}  Quantum chemistry provides a hierarchy of \emph{ab initio} methods\cite{Ochsenfeld2007} based upon Hartree-Fock, of which coupled-cluster singles and doubles with perturbative triples (CCSD(T)) is the most accurate applicable to medium-sized molecules.\cite{Bartlett2007}  Quantum Monte Carlo methods such as Green's function Monte Carlo\cite{Kalos1962,Ceperley1979} (GFMC), the closely related diffusion Monte Carlo\cite{Anderson1975,Anderson1976,Foulkes2001} (DMC), and auxiliary-field quantum Monte Carlo\cite{Blankenbecler1981} (AFQMC) produce accurate results via a stochastic sampling of the many-electron wave function, but none of these methods is exact: GFMC and DMC simulations of all but the smallest systems converge to the physically irrelevant many-boson ground state unless the fixed-node approximation is made;\cite{Anderson1975,Anderson1976} whilst AFQMC requires the phaseless approximation\cite{Zhang2003} in order to avoid an exponential growth in noise except in certain special cases.\cite{Wu2005}

The full configuration interaction (FCI) method\cite{Knowles1984} casts the Schr\"odinger equation as a matrix eigenvalue problem, in which the requirement that the many-electron wave function be anti-symmetric with respect to exchange of electrons is imposed by working in a space of Slater determinants formed from a finite basis set of single-particle wave functions.   The lowest eigenvalue and eigenvector of the FCI Hamiltonian matrix give the exact ground-state energy and wave function of the system, subject only to the error due to the finite basis set.  Whilst the computational cost of FCI scales factorially with system size, it nevertheless represents the holy grail of electronic structure methods.

In 2009, Booth, Thom and Alavi\cite{Booth2009} introduced a new stochastic approach in which the nodal structure of the ground-state wave function emerges spontaneously by sampling the discrete space of Slater determinants.  Their ``full configuration interaction quantum Monte Carlo'' (FCIQMC) method yields exact (i.e.\ FCI-quality) results whilst requiring a fraction of the memory of an FCI calculation using the same basis.  The memory required by FCIQMC simulations still scales factorially with system size, but the exponent appears to be substantially smaller than for FCI simulations. Moreover, unlike the iterative diagonaliation schemes required for FCI, the FCIQMC algorithm is readily parallelizable and can run efficiently on thousands of cores.  Alavi and co-workers have used FCIQMC to reproduce essentially every molecular FCI calculation ever done and have obtained ground-state energies for systems with Hilbert spaces many orders of magnitude larger than the largest FCI calculations.\cite{Booth2009,Booth2010,Cleland2010}  We believe that the FCIQMC method will become increasingly important in the electronic structure community, especially if improved algorithms or substantially cheaper approximations can be found.  Our motivation for exploring the behavior of the method is to provide insight into possible improvements.

% Sign problem
%  -> CH4 vs N2 vs Hubbard.
%  -> What makes problems hard?
An FCI calculation based on iterative diagonalization (using, for example, the Davidson method) requires the storage of at least two vectors, each of length equal to the size of the Hilbert space.  An FCIQMC simulation using the same basis requires the storage of the labels of the determinants occupied by a population of stochastic walkers (which, following Anderson,\cite{Anderson1975} we call ``psi-particles'' or \emph{psips}) scattered over the same Hilbert space. In order for FCIQMC to be more efficient than FCI, the number of psips required must be a small fraction of the size of the Hilbert space.  The fraction required is system dependent and provides a measure of how ``hard'' it is for FCIQMC to treat that system.  The hardness is surprisingly difficult to predict.  FCIQMC is wildly successful for some systems, such as the neon atom, where it requires only $\sim$$0.01\%$ of the memory of a conventional FCI calculation.\cite{Booth2009}  Even for the nitrogen molecule, a classic example of a strongly correlated system and a tough test for quantum chemical methods, FCIQMC used only a quarter of the memory of the equivalent FCI calculation.\cite{Booth2009}  However, FCIQMC struggles to describe the methane molecule,\cite{Booth2009} for which Hartree-Fock is a very good approximation.  We show in Sec.\ \ref{sec:fciqmc_method} that FCIQMC also struggles when applied to the Hubbard model and cannot treat systems larger than existing FCI methods unless $U$ is very small.  What is it that makes a system difficult?  Evidently the answer is more complicated than whether or not the system is strongly correlated.

The aim of this paper is to understand the FCIQMC algorithm better.  Why are some systems more difficult to treat than others?  What determines the characteristic population dynamics observed in FCIQMC simulations?  How does the cancelation of positive and negative psips ensure convergence to the many-fermion ground state?  How many psips are required to obtain correct results?  What goes wrong if the population of psips is too small?  We provide at least partial answers to all of these questions.

Section \ref{sec:fciqmc_method} reviews the FCIQMC method and applies it to the Hubbard model as an example.  In \sref{sec:sign_problem} we discuss the nature of the sign problem in FCIQMC and explain the effect of canceling positive and negative psips. Section \ref{sec:pop_dynamics} shows how our analysis of the sign problem also explains the characteristic population dynamics observed in FCIQMC simulations.  Section \ref{sec:sign_problem_free} describes several special cases in which the sign problem can be removed entirely. We offer some concluding remarks in \sref{sec:discussion}.

\section{FCIQMC method} \label{sec:fciqmc_method}

We briefly summarize the FCIQMC method here, largely following the derivations given by Booth \emph{et al.},\cite{Booth2009} with attention paid to details relevant later in this paper.

% Slater determinant space
Consider an orthonormal set of $2M$ single-particle spin-orbitals, $\{\phi_1,\phi_2,\ldots,\phi_{2M}\}$.  Previous FCIQMC work\cite{Booth2009,Booth2010,Cleland2010} has used a basis of Hartree-Fock spin-orbitals, which is often a sensible choice but not required.  One can construct an $N$-electron Slater determinant, $D_{\bi}$, by selecting any $N$ spin-orbitals (assuming $N\le2M$):
\begin{align}
D_{\bi} &= D_{i_1,i_2,\ldots,i_N} \\
       &= \frac{1}{\sqrt{N!}}
                \begin{vmatrix}
                        \phi_{i_1}(1) & \phi_{i_1}(2) & \cdots & \phi_{i_1}(N) \\
                        \phi_{i_2}(1) & \phi_{i_2}(2) & \cdots & \phi_{i_2}(N) \\
                          \vdots      &   \vdots      & \ddots &   \vdots      \\
                        \phi_{i_N}(1) & \phi_{i_N}(2) & \cdots & \phi_{i_N}(N) 
                \end{vmatrix}.
\end{align}
To ensure that all such determinants are unique (up to a sign), it is convenient to require that $i_1 < i_2 < \cdots < i_N$.

% FCI wave function
The ground-state wave function, $\Psi_0$, may be defined variationally as the $N$-electron wave function that minimizes the energy expectation value,
\begin{equation}
E[\Psi] = \bra \Psi|\hat{H}|\Psi \ket,
\end{equation}
subject to the normalization constraint $\bra\Psi|\Psi\ket = 1$.  If we restrict the search for $\Psi_0$ to the subset of wave functions that can be expanded in the basis of $^{2M}C_N$ determinants,
\begin{equation}
\Psi_0 = \sum_{\bi} c_{\bi} D_{\bi},
\end{equation}
and impose the normalization constraint using a Lagrange multiplier, the optimal coefficients $c_{0\bi}$ form the lowest eigenvector $\bc_0$ of the matrix-eigenvalue problem
\begin{equation}
\sum_{\bj} \bra D_{\bi}|\hat{H}|D_{\bj} \ket c_{0\bj} = \sum_{\bj} H_{\bi\bj} c_{0\bj} = E_0 c_{0\bi},
\end{equation}
where $E_0$ is both the Lagrange multiplier and the ground-state energy eigenvalue.  This approach yields the FCI wave function for the finite basis set used.

% imaginary-time Schroedinger equation
%  -> long-time limit
%  -> coupled diffusion equations
One way to find the ground-state eigenvector $\bc_0$ of the Hamiltonian matrix $\bH$ is to solve the imaginary-time Schr\"odinger equation:
\begin{equation}
\dd{c_{\bi}}{\tau} = - \sum_{\bj} H_{\bi\bj} c_{\bj}.
\end{equation}
The solution vector $\bc(\tau)$ converges to $\bc_0$ as $\tau \rightarrow \infty$ whenever the starting vector $\bc(\tau$$=$$0)$ has a non-zero overlap with $\bc_0$; indeed, this is also the driving principle behind other projector methods such as DMC and GFMC. The convergence is easy to demonstrate by considering the formal solution of the imaginary-time Schr\"odinger equation:
\begin{equation}
\bc(\tau) = e^{-\bH \tau} \bc(0).
\end{equation}
If the initial vector, $\bc(0)$, is expanded in terms of the complete orthonormal set of the $^{2M}C_N$ eigenvectors, $\{\bc_\alpha\}$, of the Hamiltonian matrix, $\bH$,
\begin{equation}
\bc(0) = \sum_\alpha v_\alpha(0) \bc_\alpha,
\end{equation}
the solution of the  imaginary-time Schr\"odinger equation can be written as
\begin{equation}
\bc(\tau) = e^{-\bH \tau} \sum_\alpha v_\alpha(0) \bc_\alpha = \sum_\alpha v_\alpha(0) e^{-E_\alpha \tau} \bc_\alpha.
\end{equation}
The summation is dominated by the ground-state contribution in the limit $\tau \rightarrow \infty$ and thus
\begin{equation}
\bc(\tau \rightarrow \infty) \approx v_0(0) e^{-E_0 \tau} \bc_0.
\end{equation}
The steady change in normalization due to the $e^{-E_0 \tau}$ factor is awkward but can be removed by choosing the zero of energy such that $E_0=0$.  In practice, $E_0$ is not known until the end of the simulation, so we instead solve
\begin{equation}
\dd{c_{\bi}}{\tau} = - \sum_{\bj} (H_{\bi\bj} - S\delta_{\bi\bj}) c_{\bj},
\end{equation}
where the energy shift, $S$, is adjusted slowly to keep the normalization more or less constant.  The energy shift converges to the ground-state eigenvalue, $E_0$, in the long-time limit.  Note that previous work\cite{Booth2009,Booth2010,Cleland2010} also subtracted the Hartree-Fock energy, $\EHF$, from the diagonal elements of the Hamiltonian, but this amounts merely to a redefinition of $S$. To make the subsequent notation as simple as possible, we define a ``transition matrix'', $\bT$, via
\begin{equation}
\bT = -(\bH - S\bI).
\end{equation}
Note that the minimal eigenvalue of $\bH$ corresponds to the maximal eigenvalue of $\bT$.

% algorithm
%  -> behaviour
%  -> projected energy estimator
%  -> benefit of working in a Slater determinant basis
The FCIQMC algorithm proposed by Booth \emph{et al.}\cite{Booth2009} may be summarized as follows.  Consider a collection of markers distributed over the space of determinants, $\{D_{\bi}\}$.  The markers do not move through the Hilbert space, so following Anderson\cite{Anderson1975} we call them ``psi particles'' or \emph{psips} instead of ``walkers''.  (It is easy to devise alternative FCIQMC algorithms in which the psips do move and behave much like walkers in DMC,\cite{Foulkes2001} but we use the approach of Booth, Thom and Alavi here.) Each psip has both a location, $\bi$, and a ``charge'', $q = \pm 1$, associated with it.

In one time step, $\Delta \tau$, we loop over the population of psips and allow each to ``spawn'' children (new psips) located on new determinants (which may be the same as the parent's determinant) according to the following rules:
\begin{itemize}
\item The probability that a psip at $\bi$ successfully spawns a child at $\bj$ is $|T_{\bj\bi}|\Delta \tau$.
\item If the spawning event is successful, the child psip has charge $q_{\mathrm{child}} = \sign(T_{\bj\bi}) q_{\mathrm{parent}}$, where $q_{\mathrm{parent}}$ is the charge of the parent.
\end{itemize}
At the end of the time step, pairs of psips of opposite charge on the same determinant cancel each other out (``annihilate'') and are removed from the simulation.  For example, psips on determinant $D_\bi$ with negative diagonal elements $T_{\bi\bi}$ may generate children which immediately annihilate with the parent psips.  Thus, although psips on different determinants may have different signs, all psips on any given determinant have the same sign at the end of every time step.  Any psips which remain (whether originally ``child'' or ``parent'') are merged and are allowed to spawn new psips in subsequent time steps.

The FCIQMC algorithm\cite{Booth2009} actually yields a stochastic sampling of the solution of the iterative equation
\begin{equation}
c_{\bi}(\tau + \Delta\tau) = \sum_{\bj} ( \delta_{\bi\bj} + T_{\bi\bj} \Delta \tau) c_{\bj}(\tau),
\end{equation}
which may be regarded as a finite-difference approximation to the imaginary-time Schr\"{o}dinger equation
\begin{equation}
\dd{c_{\bi}}{\tau} = \sum_{\bj} T_{\bi\bj} c_{\bj}.
\end{equation}
Assuming that $S = E_0$, as is the case once the simulation has converged on the ground state, the dominant eigenvectors of $\delta_{\bi\bj} + T_{\bi\bj} \Delta \tau$ and $T_{\bi\bj}$ are identical if $\Delta \tau ( E_{\rm max} - E_0 ) < 2$, where $E_{\rm max}$ is the largest eigenvalue of the FCI Hamiltonian matrix $\bH$. This means that there is no time-step error if $\Delta \tau$ is small enough.  In order for there to be a finite time step, the Hamiltonian must be bounded from both above and below.  It also follows that the time step is limited if high-energy states are included in the basis; for example increasing the basis set in calculations of quantum chemical systems requires a decrease in the time step.

Rather than considering all possible spawning events, it is computationally efficient to allow a psip at $D_{\bi}$ to attempt to spawn only onto its own determinant and a single connected determinant $D_{\bj}$ per time step. (Connected determinants are linked by non-zero transition matrix elements $T_{\bj\bi}$.) The connected determinant is chosen stochastically according to some \emph{generation probability} $p_{\text{gen}}(\bj \leftarrow \bi)$, which must be non-zero whenever $T_{\bj\bi}$ is non-zero and normalized such that $\sum_{\bj} p_{\text{gen}}(\bj \leftarrow \bi) = 1$. To maximise the probability of successful spawning, the generation probabilities should be as similar to $|T_{\bj\bi}|/\sum_{\bj}|T_{\bj\bi}|$ as possible; in practice, we go for speed and set the generation probabilities for spawning on all connected determinants equal. Once the candidate spawning location $\bj$ has been chosen, the spawning event is accepted with probability $|T_{\bj\bi}|\Delta \tau / p_\text{gen}(\bj \leftarrow \bi)$.\cite{Booth2009}  The effective size of the Hilbert space can be reduced by only selecting candidate spawning locations of some desired symmetry.

\begin{figure}
\includegraphics{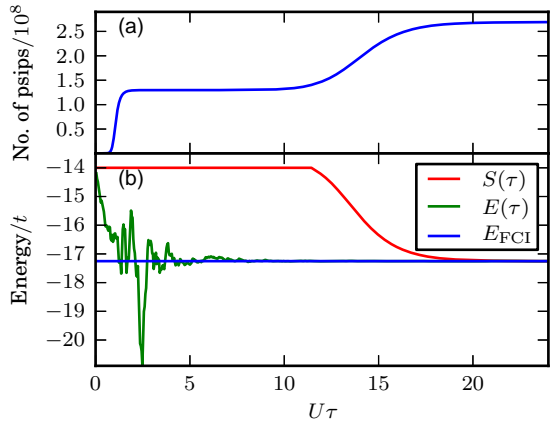}
\caption{Population dynamics and energy estimators for the 2D 18-site square lattice half-filled Hubbard model at $U=4t$.  Periodic boundary conditions are applied to a 45$^\circ$ tilted square cell with sides $3\sqrt{2}$.  The ground-state wave function has momentum $\bk=(0,0)$ and the Hilbert space formed from all determinants of this momentum contains $1.3\times10^8$ states. Graph (a) shows how the psip population evolves with simulation time.  The psip population initially grows exponentially before reaching a plateau, during which the distribution of psips settles to model the FCI wave function.  The psip population then begins to grow exponentially again, reinforcing the wave function signal, at which point the shift is allowed to vary to stabilize the number of psips.  Graph (b) shows the shift and projected energy estimators as a function of simulation time.  The FCI energy is from Ref.~\onlinecite{Becca2000}.  FCIQMC requires roughly the same number of psips as the size of the Hilbert space to converge on the ground-state wave function in this case.} \label{fig:Hubbard_dynamics}
\end{figure}

The population evolution of the psips over the course of an FCIQMC simulation has some commonly observed features, as illustrated in \fref{fig:Hubbard_dynamics}; we develop simple mathematical models to understand these results in Sections \ref{sec:sign_problem} and \ref{sec:pop_dynamics}.  The energy shift $S$ is initially set to a value greater than (i.e., less negative than) the ground-state energy eigenvalue, ensuring that the largest eigenvalue of $\bT = S\bI - \bH$ is positive and hence that total psip population grows exponentially.  Once the population reaches a critical size the simulation spontaneously enters a ``plateau'' regime, where the total psip population remains stable and during which the ground-state wave function emerges.  The height of the plateau relative to the size of the space of Slater determinants provides a measure for how hard a system is to solve using FCIQMC.  After a system- and $S$-dependent waiting time, the simulation exits the plateau phase and the psip population starts to grow in an exponential fashion again, albeit at a slower rate than before.  At this point the distribution of psips is a stochastic representation of the FCI ground-state wave function.  Allowing the psip population to increase further serves only to reduce the statistical noise, so it is convenient to hold the population constant by varying the shift from this point in the simulation. The changes in shift must be carried out smoothly and slowly to avoid introducing a bias. Following Booth,\cite{Booth2009} we use the shift-update algorithm originally proposed for DMC simulations by Umrigar\cite{Umrigar1993}:
\begin{equation}
S(\tau+A\Delta\tau) = S(\tau) + \frac{\xi}{A\Delta\tau} \ln\left(\frac{N_p(\tau+A\Delta\tau)}{N_p(\tau)}\right),
\end{equation}
where $A$ is the number of iterations between which the shift is updated, $\xi$ is a damping parameter and $N_p(\tau)$ is the total psip population at time $\tau$.  In the simulations presented here we set $A=20$ and $\xi=0.1$.  Once the psip population has settled down, the expected value $\bar{q}_\bi$ of the sum of the charges of the psips on $D_\bi$ is proportional to the weight of $D_\bi$ in the exact ground-state wave function:
\begin{equation}
\lim_{\tau \rightarrow \infty} \bar{q}_\bi(\tau) \propto c_{0\bi}.
\end{equation}
Note that annihilation of psips of opposite charge does not affect $\bar{q}_\bi$.

There are two main ways of obtaining an estimate of the ground-state energy from an FCIQMC simulation. The \emph{shift estimator} is the mean value of the shift $S$ in the constant-population-mode simulation after the plateau. The \emph{projected estimator} is obtained by noting that, for any single determinant $D_\bzero$ with a non-zero ground-state component, the quantity
\begin{align}
E &= \lim_{\tau \rightarrow \infty} \frac{ \bra D_\bzero | \hat{H} | \Psi(\tau) \ket }{ \bra D_\bzero | \Psi(\tau) \ket} \\
  &= \frac{ \sum_\bj H_{\bzero\bj} \bar{q}_\bj }{ \bar{q}_\bzero }
\end{align}
tends to the ground-state energy as $\tau \rightarrow \infty$.  Since the Hamiltonian operator only links determinants differing by two or fewer excitations, the number of terms included in the sum is limited.  Note that it is important to average $q_\bj(\tau)$ and $q_\bzero(\tau)$ separately; the projected estimator and its associated error can then be found by taking the ratio of the averages and using the covariance, respectively.  Taking $D_\bzero$ to be the determinant with the largest overlap with the exact ground-state wave function minimises the relative stochastic noise in the denominator of the above equation.  Furthermore, such a determinant will typically have single and double excitations which also have significant contributions to the ground-state wave function, and hence determinants contributing to the numerator will also often have a significant population.  The determinant with greatest overlap may not necessarily be known \emph{a priori} (or even be clearly defined, as is the case in systems studied in \sref{sec:sign_problem_free}) but in practice the Hartree--Fock determinant is usually a good choice.  The choice of $D_\bzero$ can be changed during the course of a simulation if a determinant with a particularly large population is found.

We can illustrate the main features of the FCIQMC method by applying it to the Hubbard Hamiltonian\cite{Hubbard1963} for a $d$-dimensional cubic lattice,
\begin{equation}
\label{eqn:Hub_r}
\hat{H} = -t \sum_{\langle\br,\br^{\prime}\rangle,\sigma} \hat{c}^{\dagger}_{\br,\sigma} \hat{c}^{\,}_{\br^{\prime},\sigma} + U \sum_{\br} \hat{n}_{\br,\uparrow}^{\,} \hat{n}_{\br,\downarrow}^{\,} ,
\end{equation}
where $\hat{c}^{\dagger}_{\br,\sigma}$ ($\hat{c}^{\,}_{\br,\sigma}$) creates (destroys) an electron of spin $\sigma$ on site $\br$, the number operator $\hat{n}^{\,}_{\br,\sigma} = \hat{c}^{\dagger}_{\br,\sigma}\hat{c}^{\,}_{\br,\sigma}$ counts the spin-$\sigma$ electrons on site $\br$, and the summation over $\langle \br,\br'\rangle$ includes all nearest-neigbor pairs of lattice sites. Assuming a system of $M$ sites and $2M$ spin-orbitals subject to periodic boundary conditions, the Hubbard Hamiltonian may also be written in reciprocal space,
\begin{equation}
\hat{H} = \sum_{\bk,\sigma} \epsilon_{\bk}^{\,} \hat{c}^\dagger_{\bk,\sigma} \hat{c}^{\,}_{\bk,\sigma} + \frac{U}{M} \sum_{\bk_1,\bk_2,\bk_3} \hat{c}^\dagger_{\bk_1,\uparrow} \hat{c}^\dagger_{\bk_2,\downarrow} \hat{c}^{\,}_{\bk_3,\downarrow} \hat{c}^{\,}_{\bk_1+\bk_2-\bk_3,\uparrow}, \label{eqn:Hub_k}
\end{equation}
where $\hat{c}_{\bk,\sigma}^{\dagger} = \frac{1}{\sqrt{M}}\sum_{\br} \hat{c}^{\dagger}_{\br,\sigma}e^{i\bk\cdot\br}$ is the creation operator for a Bloch state of wave vector $\bk$, the sums are over the $M$ wave vectors in the first Brillouin zone, $\epsilon_{\bk} = -2t\sum_{i=1}^{d} \cos(k_ia)$ is the one-electron kinetic energy for wave vector $\bk = (k_1, k_2, \ldots, k_d)$, and $a$ is the lattice parameter (which is set to unity from now on). The combination of wave vectors $\bk_1 + \bk_2 - \bk_3$ is assumed to have been reduced into the first Brillouin zone by addition of a reciprocal lattice vector if necessary. 

Whilst the FCIQMC algorithm may be used in the real- and reciprocal-space representations, here, for illustrative purposes, we focus on the reciprocal-space formulation, as would be appropriate if $U/t$ were small.  As $U/t$ increases and so electrons become increasingly more localised, the real-space formulation becomes more appropriate.  Indeed, as we show in \sref{sec:sign_problem_free}, the real-space basis is substantially more favourable for non-zero values of $U/t$ in the one-dimensional Hubbard model.

\begin{figure*}
\includegraphics{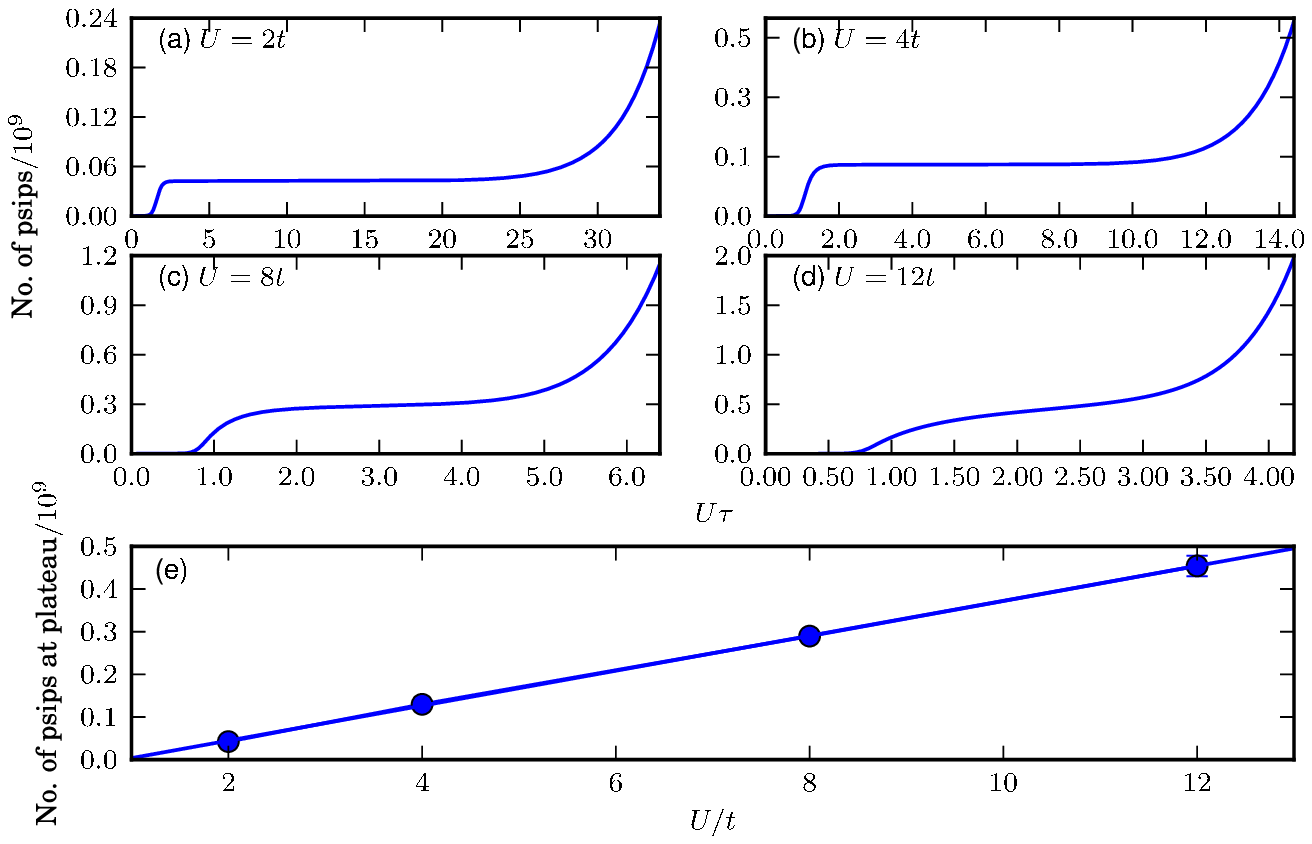}

\caption{Graphs (a)--(d) show the psip population dynamics for the 2D 18-site square lattice half-filled Hubbard model at $\bk=(0,0)$ for various values of $U/t$.  Periodic boundary conditions are applied to a 45$^\circ$ tilted square cell with sides $3\sqrt{2}$.  The data for $U=4t$ in (b) is identical to that in \fref{fig:Hubbard_dynamics}.  Graph (e) shows that the number of psips at the plateau increases linearly with $U/t$ when $U/t\geq2$. The standard errors in the numbers of psips during the plateau phases were obtained using a blocking analysis\cite{Flyvbjerg1989} and the straight line was fitted using the method of least squares as implemented in Ref.~\onlinecite{scipy}.  Unless shown, the error bars are smaller than the markers.} \label{fig:plateau_vs_U}
\end{figure*}

The imaginary-time evolution of the psip population and energy estimators during a $\bk$-space FCIQMC simulation of the 2D 18-site half-filled Hubbard model at $U/t=4$ (which is \emph{not} small) is shown in \fref{fig:Hubbard_dynamics}.  As might be expected, the $\bk$-space FCIQMC method becomes less efficient at finding the ground state of this system as $U/t$ increases --- in fact, the psip population at the plateau increases approximately linearly with $U/t$ for $U/t>2$ (\fref{fig:plateau_vs_U}).   When $U/t\geq 4$, one needs at least as many psips as there are determinants in the Hilbert space and the memory requirements are comparable to those of iterative diagonalization.  As a result, unless $U/t$ is small, $\bk$-space FCIQMC is not able to treat half-filled Hubbard model systems substantially larger than those accessible to the FCI method.

Working in a space of Slater determinants has two main advantages.  As the basis is anti-symmetric with respect to exchange of two spin-orbitals, there is no need to use the fixed-node approximation to prevent collapse to the bosonic ground state, as in DMC and GFMC. Although FCIQMC still has a sign problem (see \sref{sec:sign_problem}), the instability is not with respect to the a bosonic state and is often less severe. The second advantage is that the annihilation of psips with opposite charges proves surprisingly efficient in the discrete space of Slater determinants.  Walker cancelation can in principle cure the sign problem in continuum DMC and GFMC simulations too,\cite{Arnow1982,Carlson1985,Coker1986,Anderson1991,Anderson1992,Anderson1993,Zhang1991,Liu1994,Kalos1996,Kalos2000} but such algorithms are  substantially more complicated and less successful than the sign cancelation in FCIQMC.

\section{The origin of the sign problem in FCIQMC} \label{sec:sign_problem}

% Annihilation shown to be crucial
% Effect of removing annihilation
%  -> coupled sets of equations
%  -> lower energy 'bosonic' solution
% plots of FCIQMC with and without annihilation acting on a random matrix
Booth \emph{et al.}\cite{Booth2009} showed that annihilation is crucial in FCIQMC; without it the simulation never converges to the FCI ground state.  In this section we attempt to explain why annihilation is required and how it helps the ground state to emerge.

\begin{figure}
\includegraphics{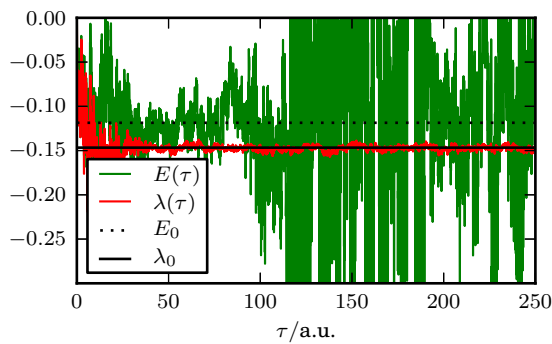}
\caption{The impact of removing annihilation from the FCIQMC algorithm applied to a $100\times100$ randomly-generated real symmetric Hamiltonian matrix $\bH$.  $E_0$ is the exact ground-state eigenvalue and $E(\tau)$ the projected estimator for $\bH$.  $\lambda_0$ and $\lambda(\tau)$ are the analogous quantities for the matrix $\bH^{+} + \bH^{-}$, where $\bH = \bH^{+} - \bH^{-}$ and all elements of $\bH^{+}$ and $\bH^{-}$ are negative (i.e., all elements of $\bT^{+}$ and $\bT^{-}$ are positive).  The simulation converges to the lowest eigenvector of $\bH^{+} + \bH^{-}$, which does not correspond to an eigenvector of the Hamiltonian matrix.}
\label{fig:no_annihilation}
\end{figure}

First, consider the effect of removing annihilation from the simulation procedure.  This means that psips of both charges are permitted to reside on and propagate from the same determinant at the same time.  We shall use $n^{+}_\bi$ and $n^{-}_\bi$ to represent the populations of positive and negative psips on determinant $D_\bi$.  It is convenient to write the transition matrix $\bT$ as $\bT^{+} - \bT^{-}$, where $\bT^{+}$ contains the positive transition matrix elements, $T^{+}_{\bi\bj} = \max(T^{\,}_{\bi\bj}, 0)$, and $\bT^{-}$ contains the absolute values of the negative elements, $T^{-}_{\bi\bj} = \max(- T^{\,}_{\bi\bj}, 0)$. All elements of $\bT^{+}$ and $\bT^{-}$ are therefore non-negative. The populations of positive and negative psips evolve according to the coupled differential equations:
\begin{equation}
\begin{aligned}
\dd{n^{+}_\bi}{\tau} &= \sum_\bj \left( T^{+}_{\bi\bj} n^{+}_\bj + T^{-}_{\bi\bj} n^{-}_\bj \right), \\
\dd{n^{-}_\bi}{\tau} &= \sum_\bj \left( T^{+}_{\bi\bj} n^{-}_\bj + T^{-}_{\bi\bj} n^{+}_\bj \right).
\end{aligned}
\label{eqn:coupled_dynamics}
\end{equation}
These can be decoupled by adding and subtracting:
\begin{equation}
\begin{aligned}
\dd{\left(n^{+}_\bi + n^{-}_\bi\right)}{\tau} &= \sum_\bj \left( T^{+}_{\bi\bj} +  T^{-}_{\bi\bj}\right) \left( n^{+}_{\bj} +  n^{-}_{\bj}\right), \\
\dd{\left(n^{+}_\bi - n^{-}_\bi\right)}{\tau} &= \sum_\bj \left( T^{+}_{\bi\bj} -  T^{-}_{\bi\bj}\right) \left( n^{+}_{\bj} -  n^{-}_{\bj}\right).
\end{aligned}
\label{eqn:decoupled_dynamics}
\end{equation}
As $\tau \rightarrow \infty$, $\bn^{+} + \bn^{-}$ tends to the eigenvector coresponding to the largest eigenvalue of $\bT^{+} + \bT^{-}$, whilst $\bn^{+} - \bn^{-}$ tends to the eigenvector corresponding to the largest eigenvalue of $\bT^{+} - \bT^{-}$.  We wish to find the latter state.  However, as explained below, the largest eigenvalue of $\bT^{+} + \bT^{-}$ is \emph{always larger} than that of  $\bT^{+} - \bT^{-}$.  Thus, the signal $\bn^{+} - \bn^{-}$ \emph{always decays} relative to $\bn^{+} + \bn^{-}$.  \fref{fig:no_annihilation} shows that performing an FCIQMC simulation without annihilation does indeed give this undesired state.  We note that a similar analysis has been performed previously for the world-line Quantum Monte Carlo method\cite{Hatano1992,Nakamura1992}.

One can prove that the largest eigenvalue of $\bT^{+} + \bT^{-}$ is always greater than or equal to the largest eigenvalue of $\bT^{+}-\bT^{-}$ as follows. Suppose that $\bc_0$ is the eigenvector corresponding to the largest eigenvalue, $\lambda_{\text{diff}}^{\text{max}}$, of $\bT^{+} - \bT^{-}$. If $\bc_0$ is normalized and the Hamiltonian matrix is real, all components of $\bc_0$ may also be chosen real and so
\begin{equation}
\label{eqn:lambdadiff}
\lambda_{\text{diff}}^{\text{max}} = \sum_{\bi,\bj} c^{\,}_{0\bi} (T^{+}_{\bi\bj} - T^{-}_{\bi\bj}) c^{\,}_{0\bj}.
\end{equation}
Now consider the vector $|\bc_0|$ with components $|c_{0\bi}|$. This vector is also normalized and may be used as a trial state in the (upside-down) variational principle for the largest eigenvalue, $\lambda_{\text{sum}}^{\text{max}}$, of $\bT^{+} + \bT^{-}$:
\begin{equation}
\label{eqn:lambdasum}
\lambda_{\text{sum}}^{\text{max}} \geq \sum_{\bi,\bj} |c^{\,}_{0\bi}| (T^{+}_{\bi\bj} + T^{-}_{\bi\bj}) |c^{\,}_{0\bj}|.
\end{equation}
The right-hand side of \eref{eqn:lambdasum} is manifestly greater than or equal to the right-hand side of \eref{eqn:lambdadiff}, so $\lambda_{\text{sum}}^{\text{max}} \geq \lambda_{\text{diff}}^{\text{max}}$.

The discrete annihilation process is difficult to model exactly\cite{annihilation_in_ode} in a differential equation.  We therefore consider a simpler annihilation process in which pairs of psips of opposite sign on the same determinant annihilate each other at a constant rate $2\kappa$, where $\kappa$ is a small positive constant and the factor of 2 is introduced solely for algebraic convenience.  This leads to the differential equations
\begin{equation}
\begin{aligned}
\dd{n^{+}_\bi}{\tau} &= \sum_\bj \left( T^{+}_{\bi\bj} n^{+}_\bj + T^{-}_{\bi\bj} n^{-}_\bj \right) - 2\kappa n^{+}_\bi  n^{-}_\bi, \\
\dd{n^{-}_\bi}{\tau} &= \sum_\bj \left( T^{+}_{\bi\bj} n^{-}_\bj + T^{-}_{\bi\bj} n^{+}_\bj \right) - 2\kappa n^{+}_\bi  n^{-}_\bi.
\end{aligned}
\label{eqn:coupled_dynamics_annihilation}
\end{equation}
The decoupled ordinary differential equations (ODEs) in \eref{eqn:decoupled_dynamics} thus become
\begin{equation}
\begin{aligned}
\dd{\left(n^{+}_\bi + n^{-}_\bi\right)}{\tau} &= \sum_\bj \left( T^{+}_{\bi\bj} +  T^{-}_{\bi\bj}\right) \left( n^{+}_{\bj} +  n^{-}_{\bj}\right) - 4\kappa n^{+}_\bi  n^{-}_\bi, \\
\dd{\left(n^{+}_\bi - n^{-}_\bi\right)}{\tau} &= \sum_\bj \left( T^{+}_{\bi\bj} -  T^{-}_{\bi\bj}\right) \left( n^{+}_{\bj} -  n^{-}_{\bj}\right).
\end{aligned}
\label{eqn:decoupled_dynamics_annihilation}
\end{equation}
It is clear how annihilation enables the FCIQMC method to converge upon the ground state of the Hamiltonian matrix: as the total psip population, $\sum_\bi (n^{+}_\bi + n^{-}_\bi)$, rises, the rate of annihilation events rises quadratically.  This causes the rate of growth of the $\bn^{+} + \bn^{-}$ signal to decrease until spawning of new psips is balanced by annihilation.  The growth of $\bn^{+} - \bn^{-}$, the desired solution, is not affected and so this signal eventually emerges.  As shown in \sref{sec:pop_dynamics}, the rates of growth of the two signals actually become the same but due to annihilation all psips on the same determinant have the same sign.
 
% Summary
To summarize, the sign problem in FCIQMC originates from the in-phase combination of positive and negative psips, which grows at a rate determined by the largest eigenvalue of $\bT^{+} + \bT^{-}$.  This eigenvalue is greater than the largest eigenvalue of $\bT^{+} - \bT^{-}$, which determines the growth rate of the physical ground state of the system. The severity of the sign problem depends upon the difference between the largest eigenvalue of $\bT^{+} + \bT^{-}$ and  the largest eigenvalue of $\bT^{+} - \bT^{-}$ --- and thus on the prevalence of negative off-diagonal elements of $\bT$ --- and upon the concentration of psips required to achieve a sufficient rate of annihilation for the growth of the in-phase signal to be suppressed.  A difficult sign problem does not necessarily imply a strongly correlated system.

% Diagonal matrix elements
\eref{eqn:coupled_dynamics} models the time dependence of the populations of positive and negative psips in a simulation in which \emph{all} annihilation is forbidden. In a real FCIQMC simulation, however, the annihilation rate per psip does not tend to zero as the psip density tends to zero. The probability of encountering and annihilating an unrelated psip vanishes, but a parent may still spawn a child of the opposite sign on its own determinant, in which case the parent and child annihilate each other. The low-density limit of the standard FCIQMC algorithm is better modeled by defining $\bT^{+}$ and $\bT^{-}$ in a different way: $\bT^{+}$ now contains \emph{all} diagonal elements of $\bT$, regardless of sign, and all positive off-diagonal matrix elements; whilst $\bT^{-}$ is zero on the diagonal but contains the absolute values of the negative off-diagonal elements. This redefinition corresponds to a change in viewpoint: instead of allowing parents to spawn children of the opposite sign on their own determinant, with subsequent annihilation, the negative diagonal elements of $\bT^{+}$ remove psips from the simulation in one step, introducing an exponential decay of the psip population on determinants for which $T^{+}_{\bi\bi} < 0$. The psip densities $n_{\bi}^{+}$ and $n_{\bi}^{-}$ remain positive at all times and the above analysis is unchanged, but the severity of the sign problem, as measured by the difference between the largest eigenvalues of $\bT^{+} + \bT^{-}$ and $\bT^{+} - \bT^{-}$, is smaller than the above analysis suggests. 

From now on it will be assumed that $\bT^{+}$ and $\bT^{-}$ are defined as explained in the previous paragraph, and hence that all diagonal elements of $\bT^{-}$ are zero.

\section{Population dynamics} \label{sec:pop_dynamics}

% Consider simplest system: a single determinant.
%  -> ODEs, introduce annihilation (somewhat crudely)
%  -> population behaviour (analytic solution)
%  -> plots
Separating the psip population into positive and negative contributions also explains the population dynamics observed during an FCIQMC simulation.  Let $\bV = \bT^{+} + \bT^{-}$, $\bp = \bn^{+} + \bn^{-}$ and $\bn = \bn^{+} - \bn^{-}$.  The decoupled ODEs in \eref{eqn:decoupled_dynamics_annihilation} can thus be written as
\begin{equation}
\begin{aligned}
\dd{p_\bi}{\tau} &= \sum_\bj V_{\bi\bj} p_\bj - \kappa (p_\bi^2 - n_\bi^2) \\
\dd{n_\bi}{\tau} &= \sum_\bj T_{\bi\bj} n_\bj.
\end{aligned}
\end{equation}
In the long-time limit, $\bn(\tau)$ becomes dominated by its ground-state component, which grows at a rate determined by the largest eigenvalue, $T_{\text{max}} = S-E_0$, of the transition matrix $\bT$:
\begin{equation}
\bn(\tau) \approx 
\alpha e^{T_{\text{max}} \tau} \bc_0 ,
\end{equation}
where $\alpha = \bc_0 \cdot \bn(0)$. Hence the evolution equation for $\bp$ becomes
\begin{equation}
\dd{p_\bi}{\tau} = \sum_\bj V_{\bi\bj} p_\bj - \kappa p_\bi^2 + \kappa \alpha^2 e^{2T_{\text{max}} \tau} c_{0\bi}^2. \label{eqn:p_diffusion}
\end{equation}

\eref{eqn:p_diffusion} is difficult to solve exactly, but the population dynamics it embodies can be illustrated using a simple one-component analogue:
\begin{equation}
\dd{p}{\tau} = V_{\text{max}} p - \kappa p^2 + \kappa n^2,
\label{eqn:Riccati_population_dynamics}
\end{equation}
where  $p(\tau)$ is the total psip population at time $\tau$, $V_{\text{max}}$ is larger than $T_{\text{max}}$, and $n(\tau) = n_0 e^{T_{\text{max}}\tau}$. It is common to start an FCIQMC simulation with a population of positive psips only, in which case the initial condition is $p(0) = n(0) = n_0$. As the initial psip population and annihilation rate are small, $p(\tau)$ grows exponentially at the start of the simulation: $p(\tau) = n_0 e^{V_{\text{max}}\tau}$. The exponential growth ceases when $p \approx V_{\text{max}}/\kappa$, at which point the $\kappa p^2$ annihilation term (which is larger than the $\kappa n^2$ term because $V_{\text{max}}>T_{\text{max}}$) counteracts it. The psip population then enters a plateau phase, remaining rougly constant until the $n(\tau)$ signal, which has been steadily growing like $n_0 e^{T_{\text{max}}\tau}$, begins to dominate. From then on the population rises exponentially again, although now at a rate determined by $T_{\text{max}}$. The ground-state wave function thus spontaneously emerges from the simulation.

The plateau begins to appear at a time $\tau_1$ determined by the equation $p(\tau_1) \approx n_0 e^{V_{\text{max}}\tau_1} \approx V_{\max}/\kappa$. Hence
\begin{equation}
\tau_1 \approx \frac{\ln(V_{\text{max}}/\kappa n_0)}{V_{\text{max}}}.
\label{eqn:tau1}
\end{equation}
More precisely, solving \eref{eqn:Riccati_population_dynamics} with the $\kappa n^2$ term omitted and assuming that $\kappa n_0/V_{\text{max}}\ll 1$ (which must be the case if the annihilation rate is negligible at the beginning of the simulation) shows that $p(\tau)$ reaches 95\% of its plateau value at time
\begin{equation}
\tau_{95\%} = \frac{\ln(19(V_{\text{max}}/\kappa n_0-1))}{V_{\text{max}}}.
\end{equation}
The end of the plateau occurs at a time $\tau_2$ determined by the equation $n(\tau_2) = n_0 e^{T_{\text{max}}\tau_2} \approx V_{\text{max}}/\kappa$, and hence
\begin{equation}
\tau_2 \approx \frac{\ln(V_{\text{max}}/\kappa n_0)}{T_{\text{max}}}.
\label{eqn:tau2}
\end{equation}
Combining Eqs.\ \ref{eqn:tau1} and \ref{eqn:tau2} yields
\begin{equation}
\frac{\tau_2}{\tau_1} \approx \frac{V_{\text{max}}}{T_{\text{max}}}.
\end{equation}

\begin{figure}
\includegraphics{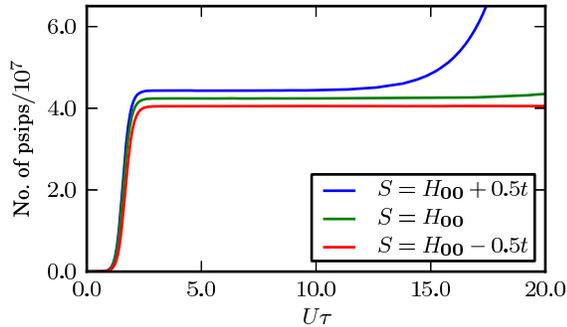}
\caption{Effect of the initial value of the shift on the plateau height in an FCIQMC simulation of the 2D 18-site square lattice half-filled Hubbard model at $U=2t$ and $\bk=(0,0)$.  Periodic boundary conditions are applied to a 45$^\circ$ tilted cell with sides $3\sqrt{2}$.  The reference determinant, $D_\bzero$, is formed from occupying the 18 lowest-energy Bloch functions.  Setting the shift slightly below $H_{\bzero\bzero}=\bra D_\bzero|\hat{H}| D_\bzero \ket$ reduces the plateau height but increases the time spent in the plateau region.}
\label{fig:vary_initial_shift}
\end{figure}

% Hubbard calculations with different initial shifts.
The energy shift $S$ appears on the diagonal of $\bT = S\bI - \bH$ and is incorporated into the diagonal elements of $\bT^{+}$; it therefore affects $\bT = \bT^{+} - \bT^{-}$ and $\bV = \bT^{+} + \bT^{-}$ equally. This dependence can be made explicit by writing $T_{\text{max}} = S + T_0$ and $V_{\text{max}} = S + V_0$, where $T_0$ and $V_0 (\geq T_0)$ are the largest eigenvalues of $\bT$ and $\bV$ when $S=0$. This leads to the equation
\begin{equation}
\frac{\tau_2}{\tau_1} \approx \frac{S+V_0}{S+T_0} = \frac{S+V_0}{S-E_0}.
\end{equation}
In most FCIQMC simulations the shift is initially held at a fixed value, typically the Hartree-Fock energy, until the desired psip population is reached. If $S$ is reduced and allowed to approach $E_0$ from above, the simulation never exits the plateau.  The duration of the plateau can be shortened by moving the initial value of $S$ further above the ground-state energy $E_0$, but only at the cost of increasing the psip population $(S + V_0)/\kappa$ at the plateau.  The effect of modifying the initial shift in an FCIQMC simulation is shown in \fref{fig:vary_initial_shift}.

\begin{figure}
\includegraphics{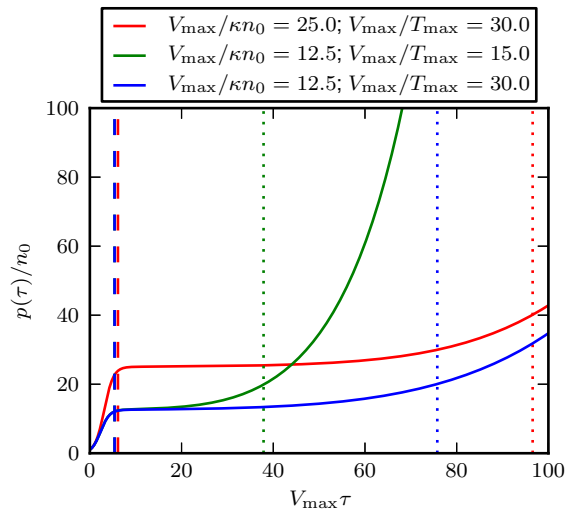}
\caption{Model population dynamics in an FCIQMC simulation.  The psip population evolves according to \eref{eqn:Riccati_population_dynamics} and shows the key features of the population dynamics in real FCIQMC calculations: an initial exponential growth phase followed by a plateau followed by a second slower exponential growth phase. As expected, the psip population reaches 95\% of its plateau value when $V_{\text{max}}\tau_{95\%} \approx \ln(19 (V_{\text{max}}/\kappa n_0-1))$, the height of the plateau is given by $p/n_0 = V_{\max}/\kappa n_0$, and the plateau ends when $V_{\text{max}}\tau_2 \approx \ln(V_{\text{max}}/\kappa n_0) V_{\text{max}}/T_{\text{max}}$.  The vertical dashed (dotted) lines show $V_{\text{max}}\tau_{95\%}$ ($V_{\text{max}}\tau_2$) for each set of parameters.  The hypergeometric function, ${_0}F_{1}$, was calculated using Ref.~\onlinecite{scipy}.}
\label{fig:pop_dynamics}
\end{figure}

The ODE in \eref{eqn:Riccati_population_dynamics} is an example of a Riccati differential equation\cite{Ince2001} and can be transformed from a quadratic first-order ODE into a linear second-order ODE using the substitution
\begin{equation}
p(\tau) = \frac{1}{\kappa u} \dd{u}{\tau}.
\end{equation}
The solution of the resultant second-order ODE is
\begin{gather}
\begin{split}
u(\tau) =& c_1 \cdot\hyperF{1-\frac{V_{\text{max}}}{2 T_{\text{max}}}}{z} \\ &+ c_2 z^{V_{\text{max}}/2T_{\text{max}}}\cdot\hyperF{1+\frac{V_{\text{max}}}{2 T_{\text{max}}}}{z},
\end{split}
\intertext{where}
z = \frac{\kappa^2 p^2(0) e^{2 T_{\text{max}} \tau}}{4 T_{\text{max}}^2},
\end{gather}
$c_1$ and $c_2$ are constants of integration, ${_0}F_{1}$ is a confluent hypergeometric limit function,\cite{Olver2010} and we have assumed that $V_{\text{max}}/(2 T_{\text{max}})$ is not an integer. The normalization of $u(\tau)$ drops out of $p(\tau) = (\kappa u)^{-1} du/d\tau$, leaving one arbitrary constant to be fixed by the initial conditions. \fref{fig:pop_dynamics} shows the resulting population dynamics for three different sets of parameters, imposing the boundary condition $p(0) = n_0$ in all cases.  Whilst this is an extremely simple model, it captures all of the features of the population dynamics seen in actual FCIQMC calculations.

Let us now return to \eref{eqn:p_diffusion} and consider what happens beyond the end of the plateau, where the psip population begins to rise again until the energy shift is adjusted and a steady state with $dp_{\bi}/dt = 0$ is attained. Since the linear $\bV$ term is negligible in comparison with the quadratic terms in this regime, \eref{eqn:p_diffusion} becomes
\begin{equation}
0 = \frac{dp_{\bi}}{dt} \approx -\kappa p_{\bi}^2 + \kappa n_{\bi}^2 ,
\end{equation}
implying that $p_{\bi} \approx \pm n_{\bi}$ and hence that $n_{\bi}^{-} \approx 0$ or $n_{\bi}^{+} \approx 0$. Determinants on which the ground-state amplitude $c_{0\bi}$ is positive are occupied only by positive psips and determinants on which $c_{0\bi}$ is negative are occupied only by negative psips. The signed psip density $n_{\bi} = n_{\bi}^{+} - n_{\bi}^{-}$ is proportional to $c_{0\bi}$ and the unsigned density $p_{\bi} = n_{\bi}^{+} + n_{\bi}^{-}$ to $|c_{0\bi}|$.

We can also use \eref{eqn:p_diffusion} to understand why the plateau psip populations plotted in \fref{fig:plateau_vs_U} are proportional to the Hubbard $U$. Assuming that the growth of the in-phase signal, $\bp$, is fast enough to allow the plateau to emerge before the ground-state signal becomes significant, the plateau occurs when
\begin{equation}
\sum_{\bi\bj} V_{\bi\bj} p_\bj \approx \kappa \sum_\bi p_\bi^2.
\end{equation}
If $U/t$ is large, the kinetic energy contributions to the Hamiltonian matrix (and hence $\bT$ and $\bV$) can be neglected and changing $U$ simply scales these matrices.  Defining $V_{\bi\bj} = U V^\prime_{\bi\bj}$ and $p_{\bi} = U p^\prime_{\bi}$ leads to the following condition for the emergence of the plateau:
\begin{equation}
\sum_{\bi\bj} V^\prime_{\bi\bj} p^\prime_\bj \approx \kappa \sum_\bi {p^\prime_\bi}^2,
\end{equation}
where $\bV^\prime$ and $\bp^\prime$ contain no dependence upon $U$.  The total psip population at the plateau, $\sum_\bi^{\,} p^{\,}_\bi = U \sum^{\,}_{\bi} p_{\bi}^{\prime}$, therefore scales linearly with $U$.

\section{Sign-problem-free systems} \label{sec:sign_problem_free}

% T & V related by a similarity transform
%   => same eigenvalues
%   => requires sign on each determinant to be well defined and consistent
%      => no annihilation
%   => no plateau
%   => thus increasing the psip population and/or propogation time serves only to reduce the stochastic error.
If there exists a similarity transform that maps $\bT$ into $\bV$, and hence makes every off-diagonal element of $\bT$ positive and every off-diagonal element of $\bH = S\bI - \bT$ negative, then $\bV$ and $\bT$ have identical eigenvalues.  Even without annihilation, the in-phase and out-of-phase signals, $\bp$ and $\bn$, grow at the same rate and there is no difficulty extracting the ground-state (out-of-phase) signal $\bn = \bn^{+} - \bn^{-}$. Such systems are sign-problem free and FCIQMC simulations of them yield correct ground-state energies with arbitrarily small psip populations, although no plateau phase is seen. Increasing the psip population serves only to reduce the stochastic error.  Indeed, by formulating the problem in the transformed basis, we can carry out an FCIQMC simulation in which no negative psips need ever appear.

A particularly simple type of similarity transformation does no more than change the signs of some of the basis determinants. No choice of signs is sufficient to render all off-diagonal elements of $T_{\bi\bj}$ positive in most cases, but there are a few interesting model systems in which simple sign-changing transformations are effective. In some models, for example, all psips spawned on any given determinant $D_{\bi}$ have the same sign regardless of the location of their parent, so that positive and negative psips never mix. A simple sign-changing transformation that multiplies every basis determinant by the sign of the psips that occupy it then makes all off-diagonal elements of $\bT$ positive.

% Heisenberg model
%  H = ... 
%  plateau a universal feature in systems with a sign problem
%  bipartite lattices are known to be free of a sign problem
%  all off-diagonal matrix elements are identical
%    => no frustration
%  1.8 million psips reproduces Runge results for the 6x6 lattice.
The antiferromagnetic Heisenberg model defined by the Hamiltonian
\begin{equation}
\hat{H} = J \sum_{\langle \br,\br' \rangle} \hat{\bS}_{\br^{\,}} \cdot \hat{\bS}_{\br^{\prime}},
\end{equation}
where $J$$>$$0$, $\hat{\bS}_{\br}$ is the vector spin operator on lattice site $\br$, and the sum runs over nearest neighbors, is such a system if the lattice is bipartite. (Note that the basis states in this example are spin configurations rather than Slater determinants.)  Indeed, the sign structure of the ground-state wave function of the bipartite Heisenberg model has long been known.\cite{Marshall1955}  Every non-zero off-diagonal matrix element of the Heisenberg Hamiltonian flips a neighboring pair of spins from down-up to up-down or vice-versa. The Hamiltonian matrix element is positive if $J>0$, implying that the corresponding matrix element of $\bT$ is negative, so the signs of children produced by off-diagonal spawning events always differ from the signs of their parents. Initially it appears as if this ought to cause a serious sign problem.  Consider, however, a bipartite system consisting of two inter-penetrating sub-lattices, arranged such that the pairs of spins flipped by off-diagonal elements of $\hat{H}$ are always on different sub-lattices. The action of any off-diagonal element of $\hat{H}$ increases the total value of $S_z$ on one sub-lattice by 1 and decreases the total value of $S_z$ on the other sub-lattice by 1. This allows all spin configurations to be assigned to one of two classes, A or B. Starting from the classical N\'{e}el state, which is arbitrarily assigned to class A, all states that differ from the N\'{e}el state by an even number of applications of $\hat{H}$ are also assigned to class A, while states that differ by an odd number of applications of $\hat{H}$ are assigned to class B.  If an FCIQMC simulation is initialized by placing a population of positive psips on the N\'{e}el state, all psips created on configurations in class A will be positive while all psips created on configurations in class B will be negative. Psips of different signs will never mix, no cancelation is required, and there is no sign problem. In fact, by applying a simple unitary transformation in which the sign of every configuration in class B is flipped, all off-diagonal matrix elements of $\bH$ can be made negative.\cite{Lieb1962}

\begin{figure}
\includegraphics{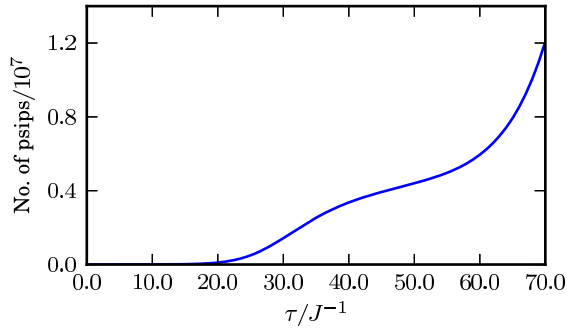}
\caption{The population dynamics of an FCIQMC simulation of the antiferromagnetic Heisenberg model for a $5\times5$ triangular lattice.  No periodic boundary conditions were imposed.  As with fermionic systems, frustrated Heisenberg models display a population plateau.}
\label{fig:frustrated_heisenberg}
\end{figure}

By way of an example, the Heisenberg model for a two-dimensional $6\times 6$ square lattice with periodic boundary conditions has a Hilbert space of $9.08\times10^9$ configurations.  Using just $1.8\times10^6$ psips we find the $E/N = -0.67886(2)J$, which is in excellent agreement with the value $E/N = -0.678871(8)J$ obtained by Runge using Green's function Monte Carlo.\cite{Runge1992}  Note that, unlike Runge, we did not use importance sampling.  FCIQMC simulations of Heisenberg models defined on other (non-bipartite or ``frustrated'')  lattices display the same behavior as for fermionic systems (\fref{fig:frustrated_heisenberg}); in particular, the population plateau is a universal feature unless the system is sign-problem free. 

% 1D Hubbard model
The Hubbard Hamiltonian, \eref{eqn:Hub_r}, is closely related to the Heisenberg Hamiltonian and is also sign-problem-free, although only in one dimension.  Consider a 1D Hubbard lattice with $N_s$ sites and $N_{\uparrow}$ ($N_{\downarrow}$) spin-up (spin-down) electrons, where the system is not necessarily half-filled and periodic boundary conditions are applied.  If $N_s$ is even and $N_{\uparrow}$ and $N_{\downarrow}$ are both odd or $N_s$ is odd and $N_{\uparrow}$ and $N_{\downarrow}$ are both even, then there exists an analagous transformation to that described above for the Heisenberg model.  (This is most easily seen if the orbitals are ordered first by spin and then by their position in the lattice.)  The ground-state energies of large 1D Hubbard models with many different choices of $N_{\uparrow}$ and $N_{\downarrow}$ can therefore be found using the FCIQMC method with very small numbers of psips.

% Unitary transforms leave eigenvalues unaltered but are not necessarily commutative.
% Hence the sign problem is not a constant for a given system, but depends upon the orbital basis.
% Plateau in the 18-site half-filled Hubbard model with Bloch orbitals removed when local orbitals are used.
This brings us on to an important point: the sign problem in FCIQMC is not constant for a given system but is dependent upon the choice of basis.  Consider two Hamiltonian matrices, $\bH_1$ and $\bH_2$, which describe the same system but have been constructed using different basis sets, one of which is obtained from the other by a transformation matrix $\bS$, so that
$\bH_1 = \bS^{-1} \bH_2 \bS$.  The corresponding transition matrices $\bT_1$ and $\bT_2$ are related in the same way. However, when we partition the two transition matrices into $\bT^{+}$ and $\bT^{-}$, the partitions are not necessarily related by the same unitary transformation. In other words, although $\bT_1 = \bS^{-1} \bT_2 \bS$, and hence $\bT_1^{+} - \bT_1^{-} = \bS^{-1} ( \bT_2^{+} - \bT_2^{-} ) \bS$,  it is not in general the case that $\bT_1^{\pm} = \bS^{-1} \bT_2^{\pm} \bS$.  Consequently, the plateau height (and thus the ease with which FCIQMC can be used to find the ground state) depends on the basis in which the Hamiltonian matrix is expressed.  

An FCIQMC simulation of the 18-site half-filled 1D Hubbard model at $U = t$
exhibits no plateau when the Hamiltonian matrix is expressed in a basis of determinants of local orbitals (as in \eref{eqn:Hub_r}), since this particular problem is sign-problem free as explained above. Yet an FCIQMC simulation of exactly the same Hamiltonian expressed in a basis of determinants of Bloch functions (as in \eref{eqn:Hub_k}) exhibits a plateau at $6.9\times10^6$ psips.  As expected, the sign problem increases the difficulty of the FCIQMC simulation in the Bloch representation. A short simulation using just $2.8\times10^5$ psips in the local orbital basis gave a ground-state energy of $-18.8423(3)t$, whereas $2.3\times10^7$ psips were required to obtain a ground-state energy of $-18.84248(8)t$ in the Bloch orbital basis.  Furthermore, our implementation currently conserves crystal momentum when using Bloch orbitals but does not make use of symmetry when using local orbitals; the Hilbert space used in the calculation with local orbitals therefore contains $2.36\times10^9$ determinants, whereas that with Bloch orbitals contains only $1.31\times10^8$ determinants.

\section{Discussion} \label{sec:discussion}

% Link
The above observations provide some degree of insight into the FCIQMC method and the factors that determine how hard it is to apply FCIQMC to any given physical system.  We now briefly comment upon other topics related to FCIQMC before summarizing our work.

% Explain success of the initiator approximation
The initiator approximation to FCIQMC (i-FCIQMC), whereby spawning events onto previously unoccupied determinants are only allowed if the parent determinant has a psip population above a specified threshold, has been shown to dramatically reduce the number of psips required to obtain excellent results in many systems.\cite{Cleland2010}  In i-FCIQMC, the effective Hilbert space grows and changes dynamically as the simulation proceeds, including only determinants that have a significant psip population or lie close (in terms of applications of the Hamiltonian) to determinants with a significant psip population.  As a result, psips are prevented from spawning in regions of low psip density and the annihilation rate is greatly increased for a given psip population.  No plateau is observed in i-FCIQMC calculations\cite{Cleland2010,Booth2010} because the growth of the in-phase combination is suppressed by annihilation even when the total psip population is low.  The i-FCIQMC approximation becomes increasingly good as the number of psips is increased because more of the Hilbert space becomes accessible and the psip dynamics more closely resembles that of the true Hamiltonian.

% Discuss CCMC
In coupled cluster Monte Carlo\cite{Thom2010} (CCMC), the excitation amplitudes of the coupled cluster wave function are stochastically sampled in manner analogous to the sampling of configuration amplitudes in FCIQMC.  Although the CCMC algorithm is much more complicated than the FCIQMC algorithm, we believe that a similar analysis holds.  The particles or ``excips''\cite{Thom_personal2011} that sample the discrete excitor space in CCMC also have positive or negative signs and annihilation is crucial for the desired (and physically meaningful) solution to emerge.  It is observed that CCMC calculations have a higher plateau than configuration interaction quantum Monte Carlo (CIQMC) calculations at the same truncation level.\cite{Thom2010} It is likely that this is because the excips have to explore a larger effective Hilbert space than the psips in the equivalent CIQMC simulation, so more excips are required to make the annihilation rate high enough to suppress the in-phase signal.

% Conclude
In summary, we have shown that the sign problem in FCIQMC simulations is caused by the growth of an unphysical dominant solution of the coupled ODE's that describe the time evolution of the densities of positive and negative psips. This unphysical solution grows faster than the ground-state solution we seek and masks the ground-state signal. A similar problem arises in DMC simulations of fermionic systems, where the dominant solution is the boson ground state. FCIQMC differs from DMC, however, in that the unphysical solution in FCIQMC is not an eigenfunction of the Hamiltonian operator.

The annihilation of psips of opposite charge suppresses the growth of the unphysical dominant solution and allows FCIQMC simulations to converge on the true ground state, but works only when a minimum (and system-dependent) threshold in the psip population is exceeded. The combination of spawning and annihilation also leads to the characteristic population dynamics observed in FCIQMC calculations.  The annihilation of psips may not be the most efficient way of suppressing the growth of the unphysical solution, and there remains a tantalising possibility that an alternative approach could lead to a superior algorithm applicable to much larger systems.

\begin{acknowledgments}
JSS is grateful for discussions on coupled cluster Monte Carlo with A.J.W.~Thom.  This work made use of the Imperial College London High Performance Computing facilities.
\end{acknowledgments}

\setlength\bibsep{2pt}

% To update the bibliography:
% delete thebibliography section 
% comment in the \bibliography line
% rebuild
% comment out the \bibliography line
% read in (include) sign_problem.bbl.
%\bibliography{sign_problem}

\begin{thebibliography}{40}
\bibitem [{ann()}]{annihilation_in_ode}%
  \BibitemOpen
  \href@noop {} {}\bibinfo {note} {{The effect can be modeled more accurately
  by using Heaviside step functions but the form shown here is more
  convenient.}}\BibitemShut {Stop}%
\end{thebibliography}


\end{document}